# Simultaneous power generation and cooling using semiconductor-sensitized thermal cells

Atsushi Hayashida[1], Hitoshi Saito[1], Yang Chunxiang[2], Taiga Nishii[2], Motokazu Ishihara[1], Yuta Nakamura[2], Kento Sunaga[2], and Sachiko Matsushita[1, 2]*

[1] elleThermo, Ltd.

INDEST 3F, Institute of Science Tokyo, Shibaura 3-3-6, Minato-ku, Tokyo 108-0023, JAPAN

[2] Institute of Science Tokyo

Mail Box: J2-48, Science Tokyo 4259 Nagatsuta-cho, Midori-ku, Yokohama 226-8503, JAPAN

**This manuscript reports a semiconductor-sensitized thermal cell (STC) that converts ambient heat into electrical power while simultaneously reducing its own temperature under isothermal conditions. Using a printable semiconductor–electrolyte architecture, we fabricate 4 cm × 4 cm devices that generate up to ~0.2 mW at 40–55 °C. During continuous discharge, the STC exhibits a transient temperature decrease followed by thermal equilibration with the environment. In contrast, periodic on–off discharge produces sustained cooling of approximately 1 °C relative to a non-discharging reference. Notably, parallel integration of four STCs yields a nonlinear enhancement of cooling (~5 °C) without a corresponding increase in electrical output. The observed behavior can be understood within a macroscopic energy-balance framework, in which time modulation of electrochemical heat consumption prevents thermal steady state. These**

**results demonstrate sustained isothermal cooling induced by heat-to-electricity conversion at practical device scales, and highlight semiconductor-sensitized thermal cells as a platform for coupled energy harvesting and thermal management.**

## Introduction

Global energy consumption has increased dramatically over the past century (Fig. 1a), reaching 592 EJ in 2024, with more than 85% still supplied by fossil fuels.[1] The International Energy Agency reports that cooling alone accounts for ~20% of electricity demand in buildings and represents one of the fastest-growing end-use loads.[2] At the device level, next-generation semiconductor chips (A10→A5 class) are projected to experience 12–15% increases in power density, leading to temperature rises approaching 9 °C.[3] Collectively, these trends highlight an urgent need for technologies that can both harvest heat and reduce thermal loads.

Isothermal power generation—energy conversion without macroscopic temperature gradients—has emerged as a promising concept. Despite the absence of an imposed temperature gradient, these devices are not perpetual-motion machines: they function as open systems sustained by a continuous inflow of environmental heat (Fig. 1b). Prior approaches include thermophotovoltaic energy conversion[4-6], plasmon-induced infrared power generation[7], near-field thermophotovoltaic devices[8,9], a metamaterial on one side of a Seebeck element[10], and an organic molecular thermoelectric system[11]. In such systems, thermal energy is transduced directly into electrical power. As a result, discharge entails a net consumption of heat, implying that a working device should exhibit a lower temperature than its non-discharging counterpart. Indeed, a previously reported metamaterial-based device exhibited a measurable temperature drop, with a discharge of ~20 µJ yielding ~0.5 K of cooling.[12]

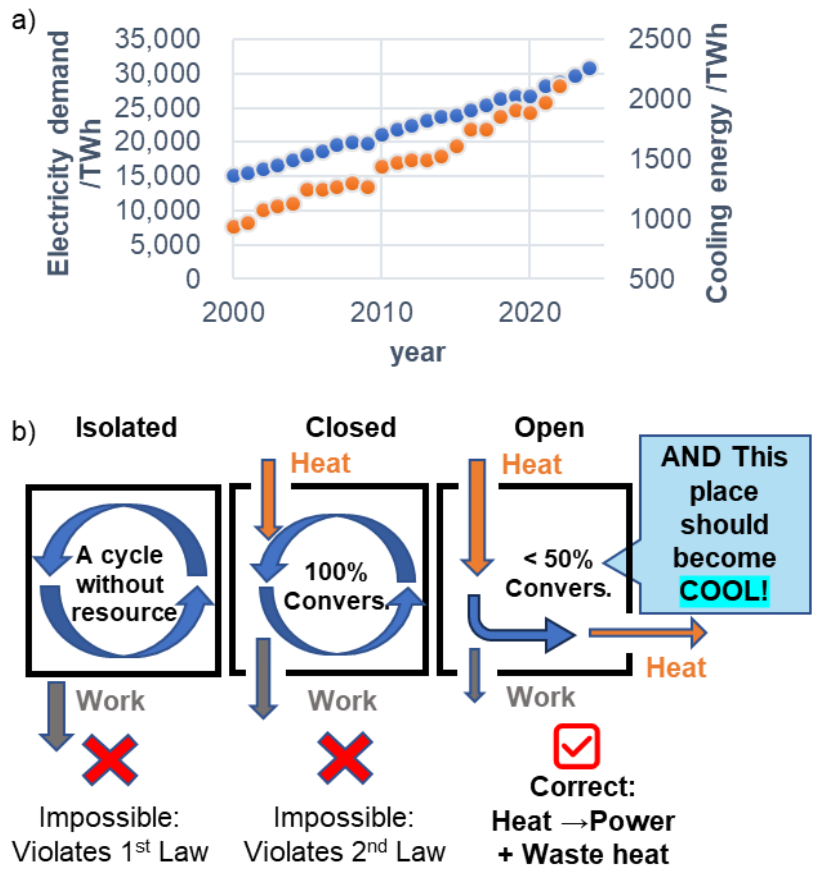

**Fig. 1 | Global energy demand and the thermodynamic basis for isothermal power generation.**

**a**, Worldwide electricity demand and cooling-related energy consumption have increased steadily over recent decades.

**b**, In an open system, thermal energy can in principle be converted into electrical power, and the power-generating site is expected to cool due to heat consumption during discharge.

Although these platforms demonstrate heat-to-electricity conversion, they typically rely on microscale devices with micro-joule-level outputs, preventing thermal effects from being observed at meaningful scales. Here we introduce a large-area semiconductor-sensitized thermal cell (STC)[13], inspired by dye-sensitized solar cells[14] but driven by thermally excited carriers[15-17] (Fig. 2a). The STC directly couples thermal excitation in a semiconductor (e.g., Germanium) with redox-ion conversion (e.g., Cu(I)/Cu(II)) in an electrolyte, enabling power generation even under isothermal conditions at surrounding temperature.[18-20] By leveraging printable semiconductor pastes and conductive-ink counter electrodes, we fabricate centimeter-scale devices capable of milliwatt-level power extraction and, critically, real-time thermal measurements. This capability allows us to quantify the interplay between electrochemical heat consumption and environmental heat inflow, revealing a cooling effect that can be sustained under periodic operation, which has not been experimentally accessible in previous isothermal power generators.

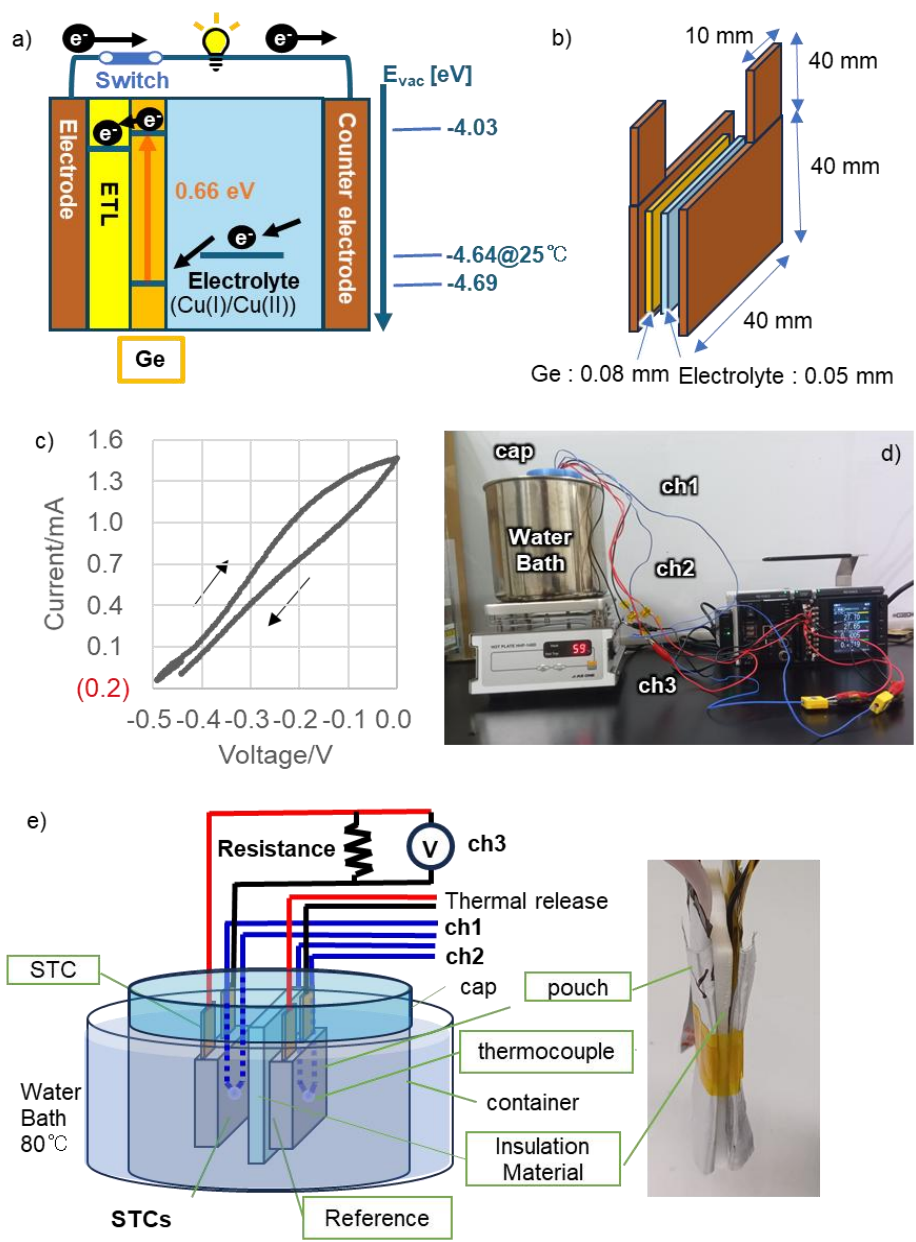

**Figure 2 | Architecture of the semiconductor-sensitized thermal cell (STC) and experimental setup for simultaneous thermal–electrical measurements.**

**a**, Band diagram of the STC illustrating thermal excitation of carriers in the semiconductor and redox-level alignment with the electrolyte. ETL, electron-transport layer.

**b**, Schematic structure of a single printable STC device.

**c**, Cyclic voltammogram of a single STC measured at 40 °C (scan rate 10 mV s$^{-1}$).

**d**, Photograph of the measurement setup used to evaluate power generation and temperature change.

**e**, Schematic of the setup shown in (d), depicting two STCs placed symmetrically inside a cylindrical chamber. Temperatures were monitored using thermocouples connected to channel 1 (ch1) and channel 2 (ch2), while discharge was controlled through channel 3 (ch3).

**Large-area STC**

Figure 2a illustrates the STC architecture. Custom printable semiconductor pastes (Supplementary Information 1) and PEG600-based CuCl/CuCl$_2$/NaCl electrolytes enabled 4 cm × 4 cm devices with a thickness of ~0.5 mm (Fig. 2b). A single device produced an open-circuit voltage of 0.44 V and a short-circuit current of 1.46 mA at 40 °C, with stable output maintained for two weeks (Fig. 2c).

**Continuous discharge: transient cooling**

To evaluate thermal behavior, an STC and a reference cell were symmetrically placed inside a cylindrical container (Methods; Fig. 2d). Under continuous discharge (981 Ω load), the STC generated 0.04 mW at room temperature, decreased to ~0.03 mW in the first 100 s due to ion depletion, then increased to ~0.10 mW as temperature rose toward 55 °C (Fig. 3a). At longer discharge durations, the electrical output gradually decreased due to concentration polarization at the electrode interface[21]; however, this trend showed no correlation with $\Delta T$ and does not affect the cooling analysis.

The temperature difference between the discharging and non-discharging devices ($\Delta T = T_{ref} - T_{power}$) reached −0.8 °C, confirming that the active STC cooled relative to the reference. $\Delta T$ subsequently relaxed toward zero as the device approached thermal equilibrium. This behaviour was reproducible across independent samples (Supplementary Information 3, samples #1–6).

The transient cooling arises when the electrochemical heat consumption $q_{out}$ briefly exceeds the environmental heat influx $q_{in}$. As the temperature difference to the surroundings diminishes, $q_{in}$ increases until it balances $q_{out}$ (dT/dt ≃ 0), at which point $\Delta T$ vanishes. A formal energy-balance derivation is provided in Supplementary Information Section 4.

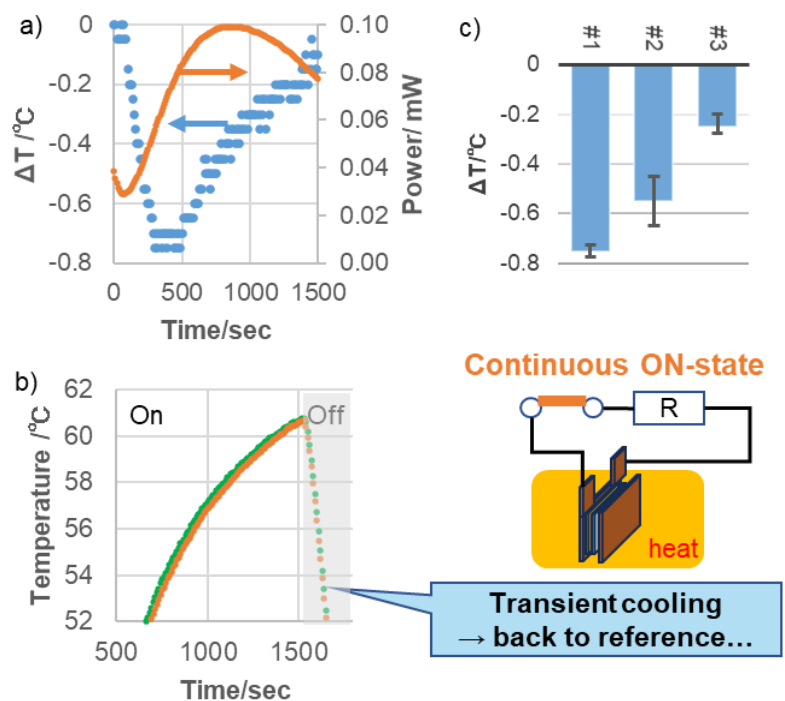

**Fig. 3 | Continuous-discharge cooling behaviour of a single STC device.**

**a**, Electrical output (orange) and temperature difference $\Delta T = T_{ref} - T_{power}$ (blue) during continuous discharge.

**b**, Absolute temperatures of the discharging STC (orange) and the reference cell (green); grey regions indicate periods when the discharge was switched off.

**c**, Maximum temperature drop for single-cell devices connected to different load resistances (981 Ω for sample #1, 461 Ω for sample #2, and 1482 Ω for sample #3). Error bars represent the measurement uncertainty of the thermal setup.

**On–off discharge for sustained cooling**

When discharge was periodically switched on/off every 120 s (Fig. 4a), the STC exhibited sustained cooling of ~1 °C relative to the reference (Fig. 4b). $\Delta T$ oscillated synchronously with the switching cycle, with a characteristic delay of ~100 s due to convective heat-transfer dynamics. This behavior was reproduced across multiple devices and cycling periods (Supplementary Information 5).

Periodic modulation of $q_{out}$ prevents the system from reaching thermal steady state: during the ON period $q_{out}$ momentarily exceeds $q_{in}$, while during the OFF period the device partially reheats. As a result, the cycle-averaged heat consumption satisfies $\langle q_{out} \rangle > \langle q_{in} \rangle$, enabling persistent negative $\Delta T$ (Supplementary Information 4). Here, $\langle \, \cdot \, \rangle$ denotes the cycle-averaged quantity. To our knowledge, this represents the first experimental demonstration of sustained isothermal cooling induced by a thermal-to-electric energy-conversion process.

A lumped thermal-resistance analysis[22,23] provides insight: with device heat capacity $C \approx 2.6$ J/K, area $A = 20$ cm², and natural-convection coefficient $h \approx 5$ W/m²K, the expected thermal time constant is $\tau \approx 260$ s, matching the experimentally observed $\Delta T$ delay (100–300 s). Thus, the temperature recovery reflects thermal equilibration with the environment. A full derivation of the internal and external thermal transport pathways is provided in Supplementary Information Section 6.

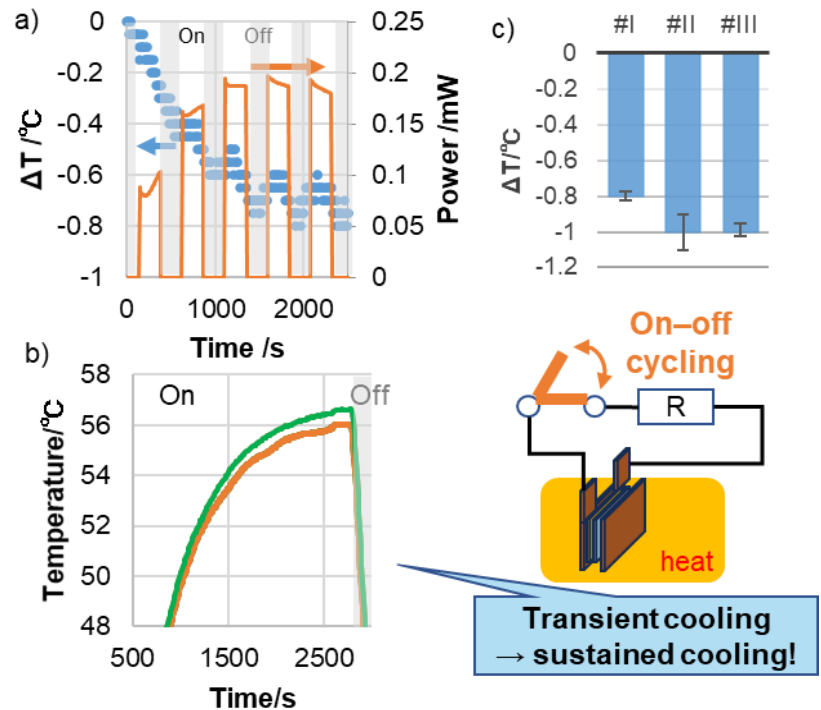

**Fig. 4 | Periodic on–off discharge of a single STC device.**

**a**, Electrical output (orange) and temperature difference $\Delta T = T_{ref} - T_{power}$ (blue) for sample #I under 200-s on/off operation. Grey regions denote intervals during which discharge was switched off.

**b**, Absolute temperatures of the discharging STC (orange) and the reference cell (green).

**c**, Peak temperature drops for samples #I, II and III, each operated with a 981 Ω load. Error bars represent the measurement uncertainty of the thermal setup.

**Arrayed devices: enhanced cooling (~5 °C)**

Strikingly, integrating four STCs in parallel yielded a temperature decrease of ~5 °C around $T_{power}=$ 55 °C (Fig. 5, Supplementary Information 3), far exceeding the 1 °C observed for a single device. Because the external load resistance was fixed, electrical power remained nearly unchanged, whereas cooling magnitude increased roughly fourfold.

This nonlinear scaling suggests that the total amount of interfacial ion reconfiguration—rather than electrical output—determines cooling strength. We hypothesize that increased electrode/electrolyte interface area amplifies entropy-related electrochemical heat consumption[24-26] and possibly ionic Peltier contributions[27-29]. The phenomenon indicates an additional heat-transport channel that is not captured by macroscopic thermal-resistance models and remains an open question for future investigation.

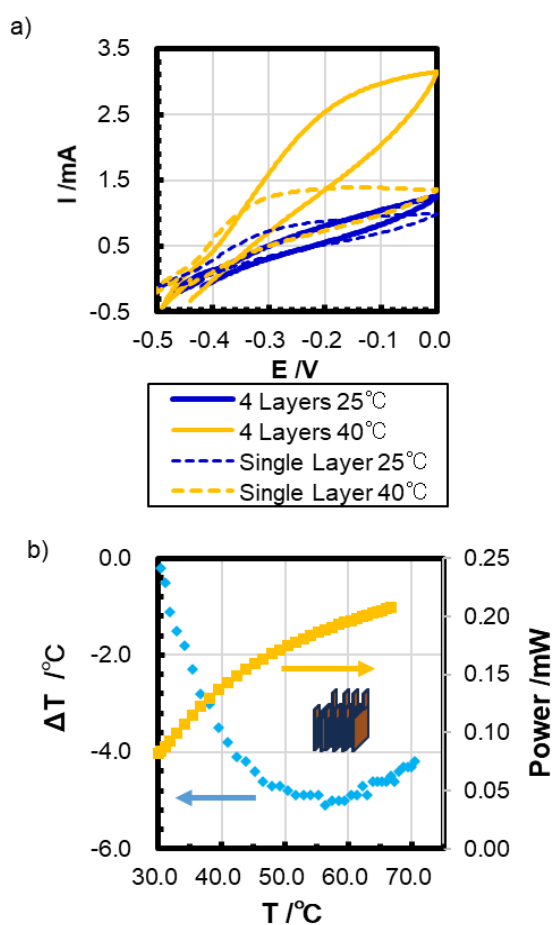

**Fig. 5 | Electrical characteristics and enhanced cooling in parallel-integrated STC arrays.**
a, Cyclic voltammograms of a single STC and a four-parallel STC array measured at 25 °C and 45 °C.
b, Temperature difference between the discharging and non-discharging sides of the four-parallel STC module under continuous discharge through a 1 kΩ external load. A marked increase in cooling is observed compared with the single-cell configuration.

**Discussion**

Our results establish semiconductor-sensitized thermal cells (STCs) as a class of energy-conversion devices capable of simultaneously harvesting heat and lowering their own temperature. A single 4 cm × 4 cm STC shows measurable cooling of approximately 1 °C under periodic on–off operation, and parallel integration of four devices produces a nonlinear enhancement, yielding a temperature drop of ~5 °C despite nearly unchanged electrical output. This behavior indicates that cooling does not scale with electrical power, but instead with the total extent of ion reconfiguration occurring during redox-driven charge transfer.

The macroscopic cooling dynamics can be understood through a heat-balance framework in which the device temperature evolves according to the competition between environmental heat influx ($q_{in}$) and electrochemical heat consumption ($q_{out}$). Under continuous discharge, $q_{out}$ initially exceeds $q_{in}$, producing transient cooling, but thermal equilibration is ultimately restored once $q_{in}$ increases sufficiently. By contrast, periodic modulation of $q_{out}$ prevents the system from reaching steady state: $q_{out}$ exceeds $q_{in}$ during the ON stage, partial reheating occurs during the OFF stage, and the cycle-averaged heat consumption $\langle q_{out} \rangle$ remains larger than $\langle q_{in} \rangle$. This mechanism enables sustained isothermal cooling, representing—to our knowledge—the first demonstration of a thermally driven electrochemical device whose operation maintains persistently negative $\Delta T$ relative to its surroundings.

Beyond this macroscopic picture, several factors likely contribute to $q_{out}$ at the microscopic level. The open-circuit voltage of the STC reflects the difference between the semiconductor Fermi level and the redox potential, consistent with thermally activated carrier excitation. During discharge, interfacial potentials shift according to the Nernst relation as redox species accumulate or deplete, and this reconfiguration of redox-active ions necessarily involves entropy changes that couple to heat consumption. Additional contributions may arise from configurational entropy at the electrolyte interface and from ionic Peltier effects associated with directed ion transport in the PEG-based electrolyte. The observed delay of ~100 s between electrical and thermal responses suggests that concentration polarization and diffusion-limited charge transfer may shape the temporal behavior of $q_{out}$. Isolating these contributions will require direct measurements of reaction entropy, ionic heat-of-transport, and redox enthalpy under controlled conditions.

These findings contrast sharply with previous demonstrations of isothermal power generation—such as metamaterial-assisted thermoelectric structures, far-infrared photovoltaic architectures, and near-field thermophotovoltaic devices—which typically operate at microjoule power levels and therefore exhibit only momentary or sub-kelvin thermal signatures. By enabling milliwatt-scale operation in a centimeter-scale device with real-time thermal monitoring, the STC platform reveals coupled thermal–electrical phenomena that were inaccessible in earlier systems.

More broadly, the ability of STCs to harvest heat while lowering their own temperature suggests a new paradigm for dual-function thermal management. Scaled deployment could enable passive cooling of high-power electronics, distributed thermal harvesting in buildings, and localized reduction of urban heat loads (Fig. 6). Advancing these applications will require deeper understanding of the cooling mechanism and systematic evaluation of large STC arrays. Nevertheless, our findings demonstrate that isothermal heat-to-electricity conversion can be amplified and utilized at practical scales, opening a pathway toward thermal-energy-positive electronics and cooling-aware energy harvesting.

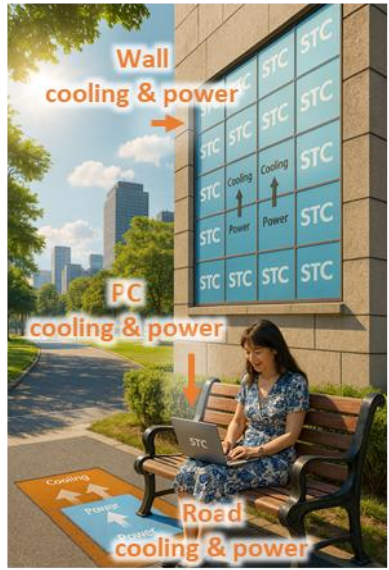

**Fig. 6 | Conceptual illustration of future applications enabled by dual-function STC cooling–energy harvesting.**

A vision of STC integration at multiple scales—from building façades and pavements to personal electronics. Embedded STC layers harvest ambient heat while locally reducing temperature, potentially lowering urban cooling loads, enabling thermally regulated high-power electronics, and contributing to sustainable energy management.

**Methods**

**Fabrication of printable semiconductor-sensitized thermal cells (STCs)**

Printable STC devices were fabricated using a custom-formulated semiconductor paste (Supplementary Information 1). The semiconductor layer was squeegee-printed onto a metal substrate and dried in vacuo at 150 °C for 1 h. A conductive-ink counter electrode was printed on a separate substrate and dried at 150 °C. The electrolyte consisted of PEG600 containing $CuCl/CuCl_2/NaCl$ (molar ratio $Cu^+/Cu^{2+}$ = 1:1, $Na^+$ = saturated)[13,30]. The two electrodes were laminated using a 0.3-mm-thick spacer with a central aperture into which the electrolyte was dispensed, defining an active area of 4 cm × 4 cm. All samples were sealed under ambient atmosphere using a heat sealer.

**Electrical measurements**

Voltage and current were measured using a Keysight 34465A digital multimeter. For discharge experiments, a fixed external resistance of 981 Ω was used unless otherwise specified. Output power was calculated as $P = V^2/R$. For parallel integration tests, four devices were connected electrically in parallel while maintaining identical load resistance.

**Thermal control and measurement setup**

To ensure symmetric heat flow into the STC and reference samples, both devices were mounted inside a cylindrical aluminum chamber (inner diameter 70 mm, height 40 mm). The two samples were affixed at equal distance from the chamber wall on a thermally insulating base plate placed at the bottom. The chamber was submerged in a temperature-controlled water bath (AS ONE, HHP-140D) to regulate ambient temperature between 25–80 °C. This configuration minimized direct heat conduction from the chamber and allowed slow, symmetric heat transfer to both samples, enabling accurate in-situ temperature measurements.

Each STC and reference device was instrumented with a type-K exposed-junction thermocouple (1 mm junction diameter, A&D), affixed directly to the electrode surface using high-temperature adhesive tape. The thermocouples were connected to a multi-channel data logger (NR-X, Keyence). Thermal-contact asymmetry was minimized by matching the lengths of the electrical measurement leads for the STC and reference, and by routing all wiring away from heat sources and electrical leads to avoid spurious thermal pickup. Verification of thermal-measurement error and channel dependence revealed that the deviation remained within ±0.4 °C across the temperature range, showing negligible dependence on ambient temperature (Supplementary Information 2).

**Temperature stabilization prior to measurement**

Before each experiment, the chamber was equilibrated in the water bath until $|\Delta T| < 0.2$ °C (typically 20–30 min). After equilibration, the bath temperature was set to 80 °C, allowing controlled heating during discharge measurements.

Strictly isothermal conditions make it difficult to resolve small heat-flow differences, so controlled heating was used to enable quantitative detection of the cooling response.

**Continuous discharge measurements**

To evaluate steady-state behavior, a single STC was connected to the 981 Ω resistor while the reference remained open-circuit. Temperature and electrical output were recorded for ~3000 s as the water-bath temperature increased from room temperature to ~60 °C (heating time ~30 min). The temperature difference was defined as

$\Delta T = T_{ref} - T_{power}$

Negative $\Delta T$ indicates cooling of the power-generating STC. This protocol was repeated across multiple devices ($n = 6$).

**Periodic on–off discharge experiments**

Sustained cooling was assessed by periodically switching the electrical load on and off using a programmable relay controller. Typical cycle durations were 120 s on / 120 s off, though shorter cycles were also tested (Supplementary Information 5). Electrical output and temperature were recorded continuously. The characteristic delay between switching and thermal response was quantified by cross-correlation analysis.

**Parallel integration tests (1-cell and 4-cell arrays)**

To investigate scaling, one or four STCs were laminated onto a common aluminum-backed support and placed in the cylindrical chamber such that all active areas received identical heat flux. Devices were connected electrically in parallel and discharged through a load. The reference device was positioned symmetrically on the opposite side of the chamber. The unusually large cooling observed for 4-cell arrays (~5 °C) was confirmed across independent samples ($n = 3$).

**Thermal modeling**

A lumped thermal-resistance model was used to estimate the thermal time constant of the device:

$\tau = C/(hA)$

The heat capacity was calculated as

$C = m\rho \approx 2.6$ J/K

using device mass $m = 2.1$ g and an effective specific heat capacity $\rho = 1.25$ kJ kg$^{-1}$ K$^{-1}$ (weighted average of PEG and metallic components).

Natural-convection heat transfer was estimated using $h = 5$ W m$^{-2}$ K$^{-1}$ and exposed surface area $A$ = 20 cm², yielding $\tau \approx 260$ s. This matches the experimental delay (100–300 s) between electrical output and thermal response.

**Reference for Methods**

**Acknowledgments**

We acknowledge the support of the following: Tohnic, Co,. and the Academy of Energy and Informatics, Science Tokyo, Japan. The Kelvin-probe microscopy and photoemission yield spectroscopy in air are measured by Riken Keiki, and SEM-EDX and laser microscopy are supported by Core Facility Center, Science Tokyo. We thank Ms. R. Kato, Science Tokyo, and Mr. Hisatsugu Yamawaki and Ms. K. Akaeda, elleThermo, Ltd. for their assistance.

**Contributions**

**S.M.** conceived and supervised the project, performed data analysis, and wrote the manuscript. **A.H.** designed and constructed the temperature-measurement system and performed the thermal experiments. **H.S.** assembled the STC devices. **C.Y.** and **T.N.** prepared the semiconductor paste and fabricated the semiconductor electrodes. **M.I.** established the STC assembly procedure. **Y.N.** measured the film thickness after paste deposition. **K.S.** measured the sheet resistance of the carbon counter electrodes.

**Ethics declarations**

The authors declare no competing interests.

**Supplementary Information is included below for completeness.**

**Supplementary Information 1. How to prepare the 4 cm-sq. STC.**

**1.1. Fabrication of the semiconductor paste**

Germanium powder (Furuuchi Chemical, ≥99.9%) was mechanically milled and sieved through a 40-µm mesh. The particle-size distribution after milling is shown in Fig. S1a. The surface oxide layer was removed by treatment in 5 wt% HF(aq) for 15 min under a fume hood, followed by repeated rinsing with deionized water (18.2 MΩ·cm) and drying under vacuum.

The optical bandgap of the resulting Ge powder was evaluated by UV–Vis–IR spectroscopy (JASCO V-770) and Tauc analysis, yielding $E_g \approx 0.66$ eV (Fig. S1a).

The surface work function was measured by ambient photoelectron spectroscopy (Riken Keiki AC-3), yielding −4.6 to −4.9 eV, with an average of −4.69 eV, as shown in Fig. 1a.

The cleaned Ge powder was mixed with a polymeric binder in a 1:1 volume ratio using an agate mortar. A volatile organic solvent was added to adjust viscosity suitable for squeegee coating. The paste was mixed over 20 min until homogeneous.

**1.2. Fabrication of the semiconductor electrode**

Metal foil (30 µm-thickness) was cut as 4 cm × 4 cm with a tab (1 cm × 4 cm). The metal substrate was coated on both sides using an adjustable-gap film applicator (gap: 200 µm) and dried at 150 °C for 60 min.

The resulting semiconductor layer thickness was approximately 80 µm (Fig. S1b). Surface roughness was examined by laser confocal microscopy (KEYENCE VK-X200), shown in Fig. S1c. The cross-sectional SEM image is shown in Fig. S1e.

**1.3. Fabrication of the counter electrode**

Conductive carbon paste was screen-printed onto identical metal foils using a 200-mesh stainless screen. After printing, the films were dried at 150 °C over 30 min., producing a carbon layer of ~90 µm thickness and a sheet resistance of ~10 Ω/□.

**1.4. Electrolyte preparation**

The electrolyte consisted of PEG1540 (FUJIFILM Wako) containing CuCl (0.25 mmol per g-PEG), $CuCl_2$ (0.25 mmol per g-PEG), and saturated NaCl. All components were mixed inside an Ar-filled glovebox ($O_2$, $H_2O$ < 1 ppm). The physicochemical characteristics of this electrolyte have been reported previously (ACS Electrochem. 2025, 1, 7, 1076–1081).

**1.5. Cell assembly**

A mesh spacer (53 µm thickness) was placed on top of the counter electrode, and 100–150 µL of the electrolyte was dispensed uniformly. The two electrodes were laminated using a 0.3-mm-thick film (Clever Co.)

with a central aperture into which the electrolyte was dispensed, defining an active area of 4 cm × 4 cm. All samples were sealed under ambient atmosphere using a heat sealer (T-130K, Fujiimpulse Co.).

**1.6. Cell characteristics.**

Electrochemical measurements were performed using a two-electrode configuration with the semiconductor electrode serving as the working electrode. All measurements were carried out using a VSP-300 potentiostat (Bio-Logic Science Instruments). CVs were recorded at 10 mV s$^{-1}$ after stabilizing the device in a temperature-controlled incubator.

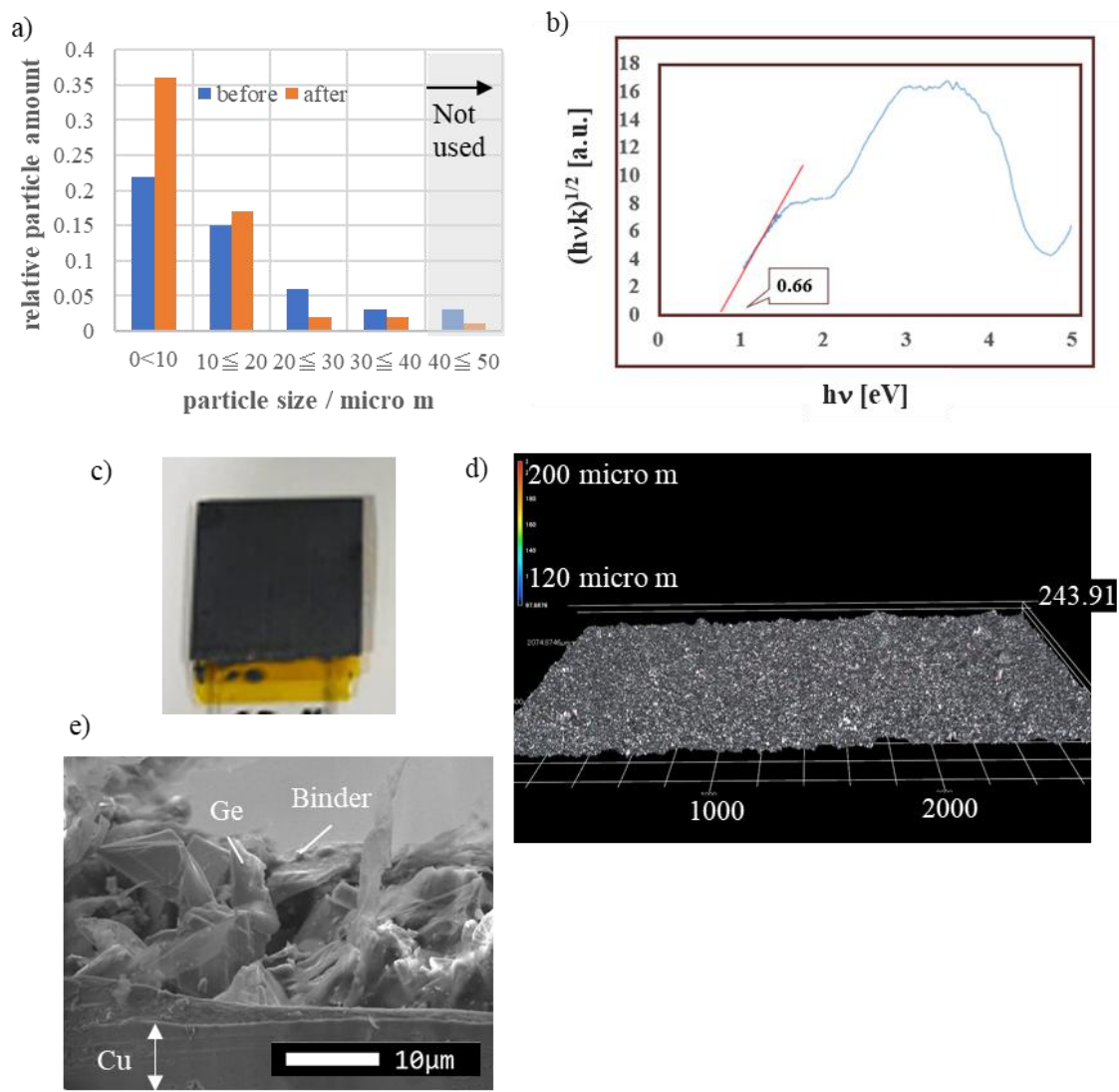

**Fig. S1 | Characterization of the Ge powder and printed semiconductor electrode.**

**a**, Particle size distribution of Ge powder before and after sieving. Particles larger than 40 μm were excluded from STC fabrication ("Not used"). The proportion of fine particles (<10 μm) increased significantly after processing. Relative particle amount of the Ge powder before and after milling.

**b**, Tauc plot of the processed Ge powder used in this study.

**c,** Photograph of the printed Ge-paste electrode (4 cm × 4 cm).

**d,** Three-dimensional surface morphology of the Ge-paste electrode obtained by laser optical microscopy.

**e,** Cross-sectional SEM image of the fabricated Ge-paste electrode.

**Supplementary Information 2. Verification of thermal-measurement error and channel dependence.**

**a,** Schematic of the electrical configuration used to evaluate measurement error. Two reference cells (Ref. A and Ref. B) were prepared and alternately connected to channel 1 and channel 2 to assess potential channel-dependent offsets. No electrolyte ions were added to the reference cells in order to eliminate effects arising from ion-related redox reactions at the semiconductor surface and entropy changes associated with interfacial ion structuring.

**b,** Temperature differences measured between Ref. A and Ref. B. The deviation remained within ±0.4 °C across the entire temperature range, indicating negligible dependence on ambient temperature. Comparable error levels were observed even after swapping the references between channels. Because the STCs are handmade, small variations in metal-foil dimensions and paste volume likely caused slight differences in heat capacity, contributing to the residual offsets.

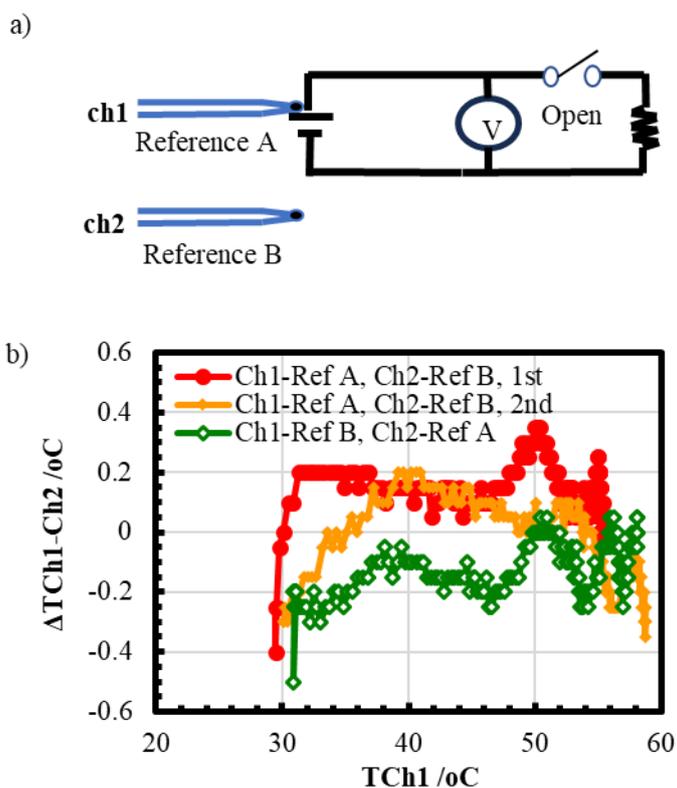

**Supplementary Information 3. Reproducibility of continuous discharge.**

We examined the reproducibility of the cooling behavior by varying both the resistance and the number of STC units. In all cases, the onset of cooling and its subsequent relaxation were consistently observed over time. Increasing the number of STC units enhanced the magnitude of the cooling effect; however, no clear correlation with the output signal was identified. We speculate that the discrepancy may arise from differences in heat capacity, although this remains to be investigated in future work.

| Sample # (Figure) | Continuous/ Interval discharge | STC | Reference | Resistance [Ω] | ΔT before meas. [°C] |
|---|---|---|---|---|---|
| #2 (a, b) | Continuous | 1 | Metal+PEG600+separator (without semiconductor and ions) | 461 | 0.2 |
| #3 (c, d) | Continuous | 1 | Metal+PEG600+separator (without semiconductor and ions) | 1482 | 0.05 |
| #4 (e, f) | Continuous | 4-parallel | Metal+PEG600+separator (without semiconductor and ions) | 981 | 0.6 |
| #5 (g, h) | Continuous | 4-parallel | Metal+PEG600+separator (without semiconductor and ions) | 6720 | -0.2 |
| #6 (i, j) | Continuous | 4-parallel | Metal+PEG600+separator (without semiconductor and ions) | 6720 | 1.1 |

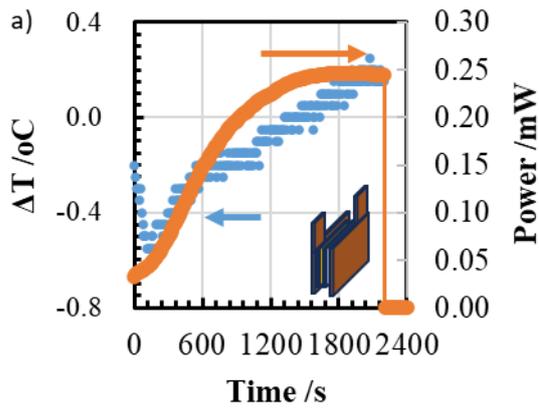 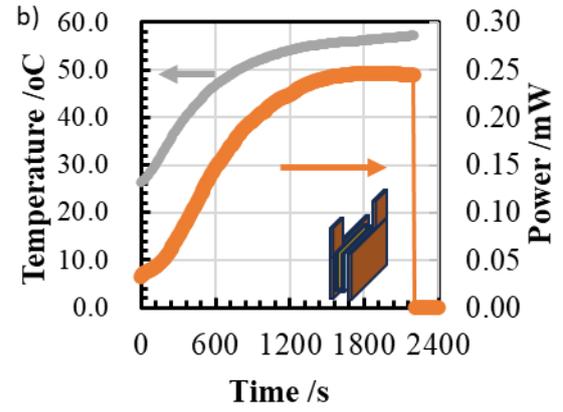
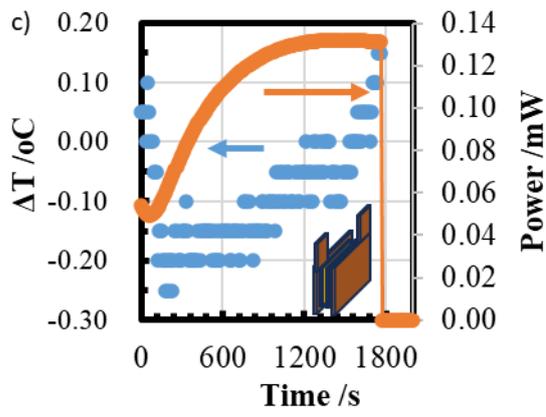 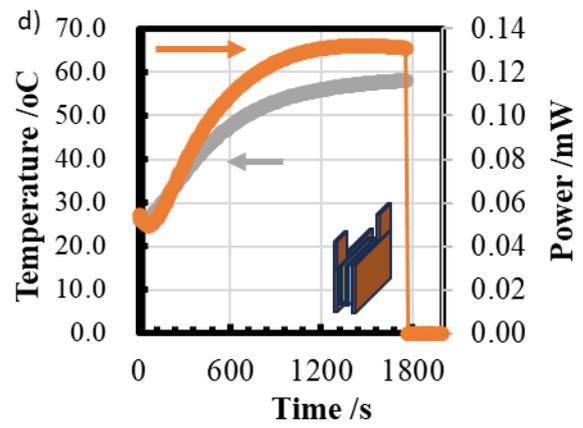

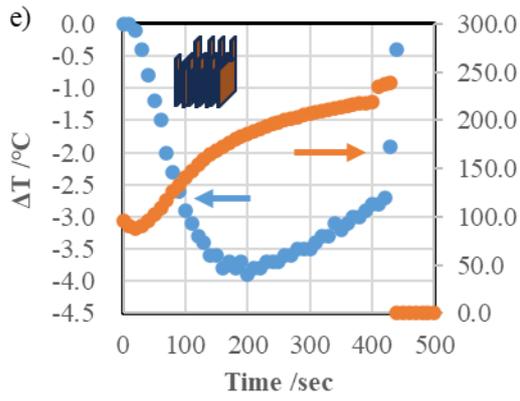
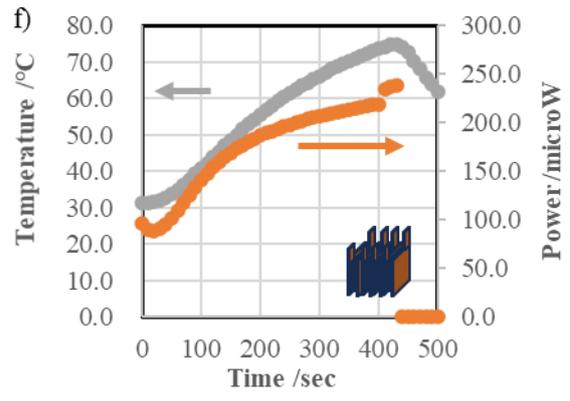
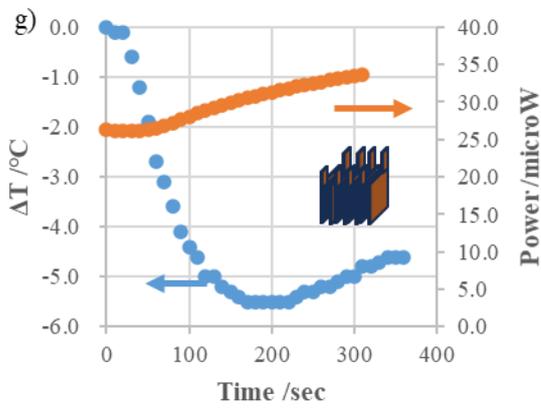
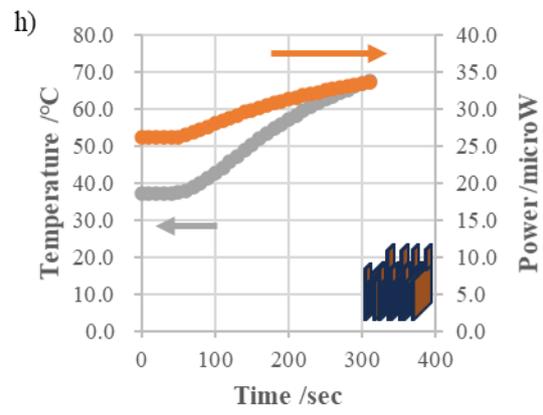
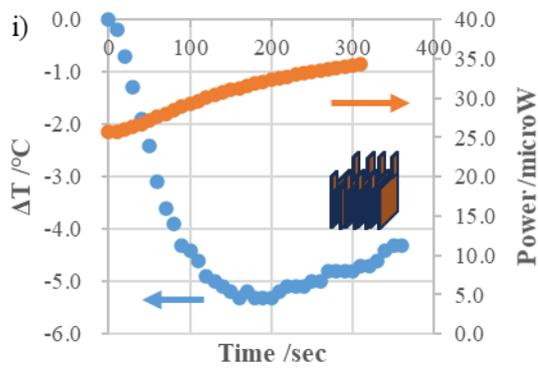
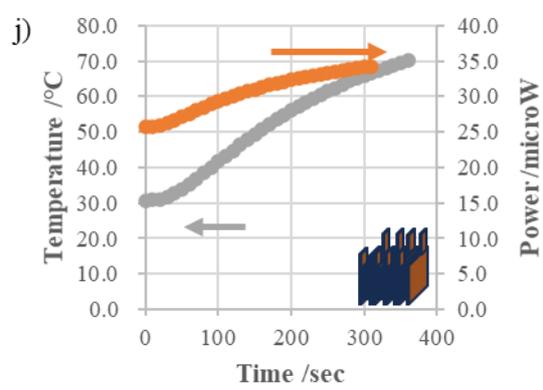

**Supplementary Information 4. Energy-balance model explaining transient and sustained cooling in STCs.**

**4.1. Governing heat-balance equation**

The temperature of the device is determined by its internal energy $Q$. The temporal evolution obeys

$$\frac{dQ}{dt} = q_{in} - q_{out}$$

where $q_{in}$ and $q_{out}$ represent the incoming and consumed heat fluxes, respectively. Using $C\,dT/dt = dQ/dt$, the temperature dynamics are

$$C\frac{dT}{dt} = q_{in} - q_{out}$$

This framework captures both the transient cooling and the recovery dynamics observed in the main text.

**4.2. Continuous discharge: approach to thermal steady state**

When discharge is sustained, $q_{out}$ remains nearly constant, whereas $q_{in}$ increases with decreasing temperature difference to the surroundings. Thus the system inevitably reaches a steady state ($q_{in} = q_{out}$, $dT/dt \approx 0$), explaining why continuous discharge cannot maintain negative $\Delta T$ indefinitely (Fig. 3).

**4.3. Periodic discharge: conditions for sustained cooling**

We now consider the case where discharge is periodically switched ON and OFF, corresponding to time-dependent $q_{out}(t)$.

• **ON period**

$q_{out}$ is large, and

$$q_{out} > q_{in}$$

So the device cools.

• **OFF period**

Electrochemical heat consumption ceases:

$$q_{out} = 0,$$

leaving only $q_{in}$, and the device reheats.

Over one full cycle, the key quantity is the cycle-averaged heat balance:

$$\langle q_{in} \rangle - \langle q_{out} \rangle.$$

From this, three regimes follow naturally:

a. $\langle q_{out} \rangle > \langle q_{in} \rangle$

    → net heat removal

    → temperature decreases over cycles

b. $\langle q_{out} \rangle = \langle q_{in} \rangle$

    → thermal balance

    → temperature stabilizes

c. $\langle q_{out} \rangle < \langle q_{in} \rangle$

    → net heat inflow

    → temperature increases

Thus, sustained cooling is possible only when the cycle-averaged heat consumption exceeds the incoming heat flux:

$$\langle q_{out} \rangle > \langle q_{in} \rangle$$

which matches the experimental observation of persistent negative $\Delta T$ under periodic discharge (Fig. 4).

### 4.4. Implications

This analysis provides the theoretical basis for the key observation in the main text: periodic modulation of discharge enables sustained isothermal cooling, whereas continuous discharge cannot. It also explains why the thermal response exhibits a characteristic delay determined by the external convective thermal resistance (see Supplementary Information Section 6).

**Supplementary Information 5. Reproducibility of interval (on–off) discharge–induced cooling in STC devices.**

Temperature difference $\Delta T = T_{ref} - T_{power}$ and electrical output measured for multiple STC samples under periodic on–off discharge. All devices exhibit sustained negative $\Delta T$ with similar phase delay, confirming that cycle-induced cooling is reproducible across samples. Error bars represent measurement uncertainty.

| Sample # (Figure) | Continuous/Interval power generation | Interval [sec] | Reference | Resistance [Ω] | ΔT before meas.[°C] |
|---|---|---|---|---|---|
| II (a, b) | On/Off | 120 | Metal+PEG600+separator + semiconductor and ions | 981 | 0.05 |
| III (c, d) | On/Off | 120 | Metal+PEG600+separator + semiconductor and ions | 981 | -0.2 |
| IV (e, f) | On/Off | 240 | Metal+PEG600+separator + semiconductor and ions | 98.4 | -0.05 |

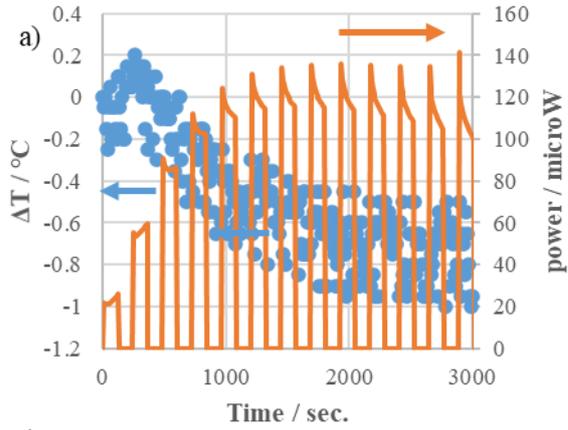 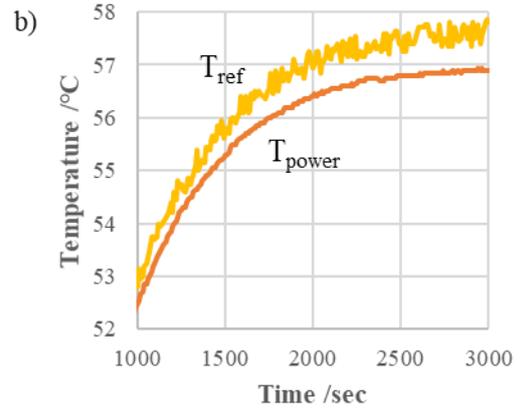
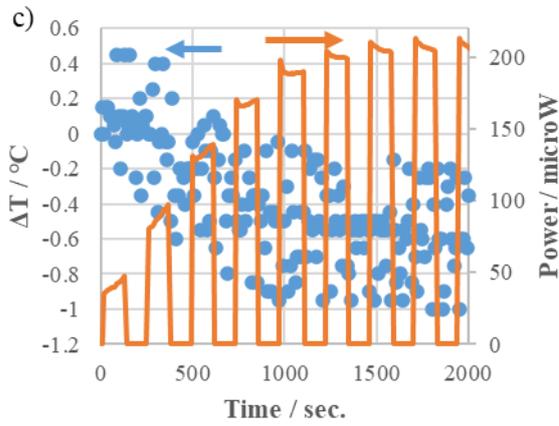 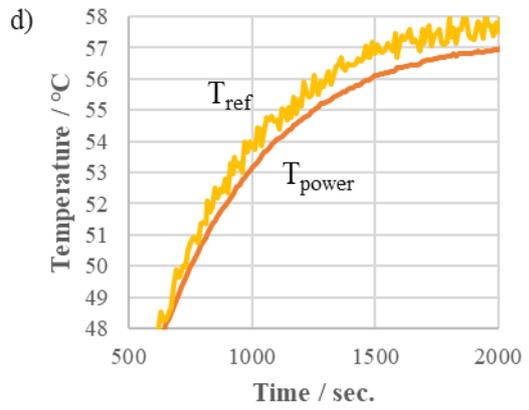
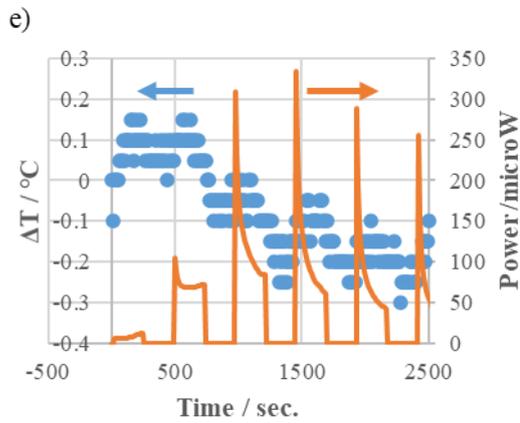 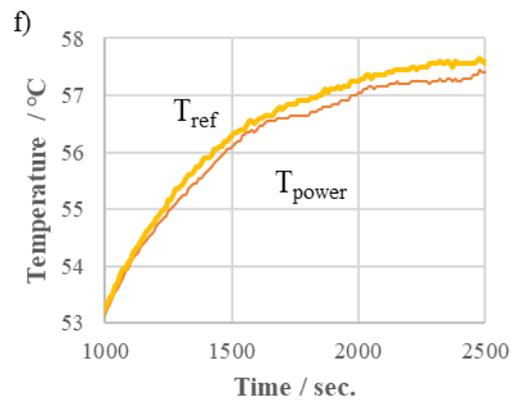

**Supplementary Information 6. Internal Thermal Transport in the STC Device.**

**6.1. Thermal conduction within the STC**

The dominant thermal pathway in our measurement system is the heat transfer across the air/Al-pouch interface. To confirm that internal heat conduction within the STC is not rate-limiting, we estimated the characteristic conduction timescales through the Al pouch and the internal device layers.

The aluminum pouch used in the experiments had a thickness of 0.11 mm, with a thermal conductivity $k \approx 0.3–1$ W m$^{-1}$ K$^{-1}$. The STC thickness was ~0.5 mm, composed primarily of electrolyte. Assuming PEG-based electrolyte properties:

- density $\rho \approx 1100$ kg m$^{-3}$
- heat capacity $C \approx 2000$ J kg$^{-1}$ K$^{-1}$
- thermal conductivity $k \approx 0.3$ W m$^{-1}$ K$^{-1}$

The thermal diffusivity is:

$$\alpha = k/(\rho C) \approx 1.4 \times 10^{-7} \text{ m}^2/\text{s}$$

Assuming heat originates at the mid-plane and propagates across half the device thickness (L/2 = 0.35 mm = 3.5 × 10$^{-4}$ m), the characteristic diffusion time is:

$$t \sim L^2/\alpha = (3.5 \times 10^{-4})^2 / (1.4 \times 10^{-7}) \approx 1.8 \text{ s}$$

Thus, internal heat conduction completes on the order of only a few seconds.

**6.2. Dominant thermal bottleneck: external convection**

Because internal conduction is fast, it does not limit the thermal response. Instead, the rate-limiting step is the external heat exchange between the Al-pouch surface and ambient air, governed by natural convection. Under these conditions, the device temperature evolves according to:

$$\frac{dT}{dt} = \frac{hA}{C}(T_{power} - T_{air})$$

where $h$ is the natural convection coefficient, $A$ is the exposed area, and $C$ is the heat capacity of the STC.

This relation forms the basis of the lumped thermal-resistance model used to estimate the thermal time constant in the main text.